\documentclass[twocolumns]{aa} 

\usepackage{graphicx}

\usepackage{listings}
\usepackage{txfonts}
\usepackage{url}
\usepackage{booktabs}
\begin{document} 
\title{Kete}
\subtitle{Predicting Known Minor Bodies in Images}

\author{D. Dahlen\inst{\ref{inst1}
,\ref{inst2}}, Y. G. Kwon\inst{\ref{inst2}}, J. R. Masiero\inst{\ref{inst2}}, T. Spahr\inst{\ref{inst3}}, A. K. Mainzer\inst{\ref{inst4}}}

\institute{Institut f{\"u}r Geophysik und Extraterrestrische Physik, Technische Universit{\"a}t Braunschweig,  Mendelssohnstr. 3, 38106 Braunschweig, Germany \\
    \email{dar.dahlen@gmail.com}\label{inst1}
\and Caltech/IPAC, 1200 E California Blvd, MC 100-22, Pasadena, CA 91125, USA\label{inst2}
\and NEO Sciences, USA\label{inst3}
\and University of California, Los Angeles, CA, USA\label{inst4}
}

\date{Submitted -- / Accepted --}

\abstract{
Kete is an open-source software package for quickly and accurately predicting the positions and magnitudes of asteroids and comets in large-scale, all-sky surveys. It can predict observable objects for any ground or space-based telescope. Kete contains a collection of tools, including simple optical and thermal modeling, $n$-body orbit calculations, and custom multi-threaded \texttt{SPICE} kernel support. It can be used for observation planning, pre-discovery of detections at a large scale, and labeling known solar system objects in images. Here we demonstrate some of the capabilities by predicting all observations of every numbered asteroid seen by the Wide-field Infrared Survey Explorer (WISE) and Zwicky Transient Facility (ZTF) surveys during single years of their operations, predicting locations and magnitudes of 756,999 asteroids in over 11 million images.
}


\titlerunning{Kete: Open-source software package for solar system small-body science}
\authorrunning{D. Dahlen et al.}

\maketitle


\section{Introduction}

Around the turn of the millennium, the advent of automated all-sky surveys marked the beginning of the data-intensive era in Solar System astronomy. Over the past two decades, ground-based surveys, such as the Catalina Sky Survey \citep{Larson1998}, the Sloan Digital Sky Survey (SDSS; \citealt{York2000,Stoughton2002}), the Lincoln Near-Earth Asteroid Research (LINEAR) program \citep{Stokes2000}, the Asteroid Terrestrial-impact Last Alert System (ATLAS; \citealt{Tonry2018}), the Panoramic Survey Telescope and Rapid Response System (Pan-STARRS; \citealt{Kaiser2002}), and the Zwicky Transient Facility (ZTF; \citealt{Masci2018}), and space-based missions such as the Wide-field Infrared Survey Explorer (WISE; \citealt{wise}) and its planetary science extension, NEOWISE \citep{Mainzer2011}, have substantially advanced the field.

Recent and upcoming surveys, notably the Legacy Survey of Space and Time (LSST) conducted by the Vera C. Rubin Observatory \citep{Jones2020}, the Spectro-Photometer for the History of the Universe, Epoch of Reionization, and Ices Explorer (SPHEREx; \citealt{Crill2020}), the Gaia mission \citep{gaia}, and Near Earth Object Surveyor (NEOS;  \citealt{neos}), are expected to expand small-body datasets by several orders of magnitude. Using these resources fully requires robust and adaptable computational infrastructures capable of accurate, high-throughput analysis across heterogeneous data products. Such tools will not only streamline scientific workflows but also enhance the interpretability and contextualization of small-body populations.

Motivated by this need, \texttt{Kete} is designed as a versatile framework for small-body data analysis. Its central functionality is to identify known Solar System objects detectable by any ground- or space-based observatory, enabling accurate simulations of survey performance and object cross-identification in both archival and real-time survey imagery.

\texttt{Kete} consists of individual components, including representations of Field of Views (FoV) of telescopes (Section \ref{geometry}), orbit propagation to compute ephemerides and sky positions for Solar System objects (Section \ref{prop}), and flux estimation to predicts visible and infrared flux densities or magnitudes (Section \ref{flux}). These components can be composed together to achieve \texttt{Kete}'s core objective: determining which asteroids and comets fall within a given telescope footprint. The framework also supports broader use cases, including observation planning, modeling of cometary dust tails, and analysis of orbital evolution during close planetary encounters, as described in later sections.

\texttt{Kete} is implemented in Python with a compiled backend in Rust. This hybrid design, similar to performance-oriented packages such as \texttt{NumPy} \citep{Harris2020}, allows computationally demanding operations while maintaining a user-friendly interface. Rust was selected for its runtime performance comparable to C, multi-core support, and its strong guarantees around memory safety. The source code is publicly available at \url{https://github.com/dahlend/kete}. Online documentation along with details on how to install it\footnote{\texttt{Kete} is installable via \texttt{pip install kete} and supports all current Python versions (v3.9 or later).} and numerous worked examples (including how to reproduce several of the figures presented) may be found at \url{https://dahlend.github.io/kete/}.

\begin{figure*}[htb]
\centering
\includegraphics[width=0.9\textwidth]{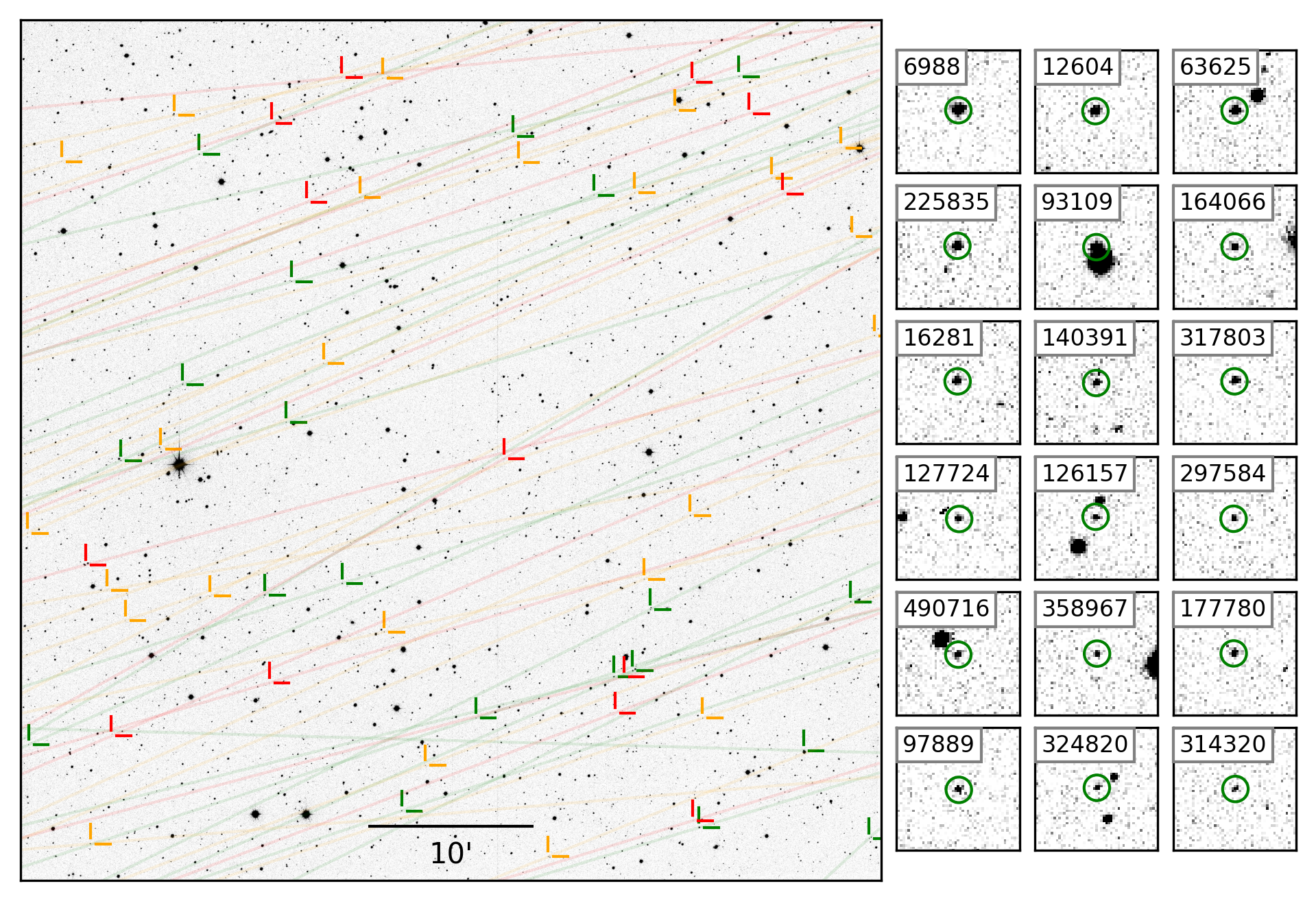}
\caption{Example ZTF image taken on 2021-10-25 at 06:11:13.935 UT with the $zi$ filter, containing 57 numbered asteroids. The left panel shows the full image, with positions of the asteroids labeled using \texttt{Kete}. Color coding indicates predicted $V$-band magnitudes: green for $V < 19.5$, red for $V > 20.5$, and orange for intermediate values. Faint lines indicate the orbit direction of each asteroid in the image. The right panel displays cutouts of the brightest asteroids, with circles marking the predicted positions and labels showing the corresponding asteroid numbers. This image represents one of the densest examples from the simulation described in the text, drawn from a total of 9.976 million ZTF images analyzed.}
\label{fig:ztf_frame}
\end{figure*}

\section{Simulation Description}

The primary goal of \texttt{Kete} is to predict the visibility and apparent positions of known asteroids and comets within telescope images. To demonstrate its utility, we performed two simulations that computed the positions, velocities, and apparent magnitudes of all numbered asteroids observed by ZTF (optical, ground-based) and WISE/NEOWISE (infrared or IR, space-based) telescopes, described in \cite{ztf}, and \cite{wise}, respectively. These two telescopes were chosen as they provide examples of both a ground- and space-based telescopes, examples of visible and IR band observations, and have publicly available metadata describing every image captured during their operations, thus demonstrating the broad applicability of \texttt{Kete}. These simulations involved computing the ephemerides of every numbered asteroid across all images acquired by both telescopes over the course of a year. Both simulations were performed on an Apple M3 laptop, each taking less than an hour to complete. The ZTF simulation was run for the year 2021, with a total of 9,976,181 images checked if they contained any of the 756,999 numbered asteroids available at the time of simulation. Figure~\ref{fig:ztf_frame} shows one of the ``busiest'' images for that year of ZTF, containing 57 asteroids geometrically located within the single image FoV (0.75 degrees square). Note that although these objects fall within the imaged area, their geometric presence does not guarantee photometric detectability; many are fainter than the telescope’s limiting magnitude.

\begin{figure*}[htb]
\centering
\includegraphics[width=0.87\textwidth]{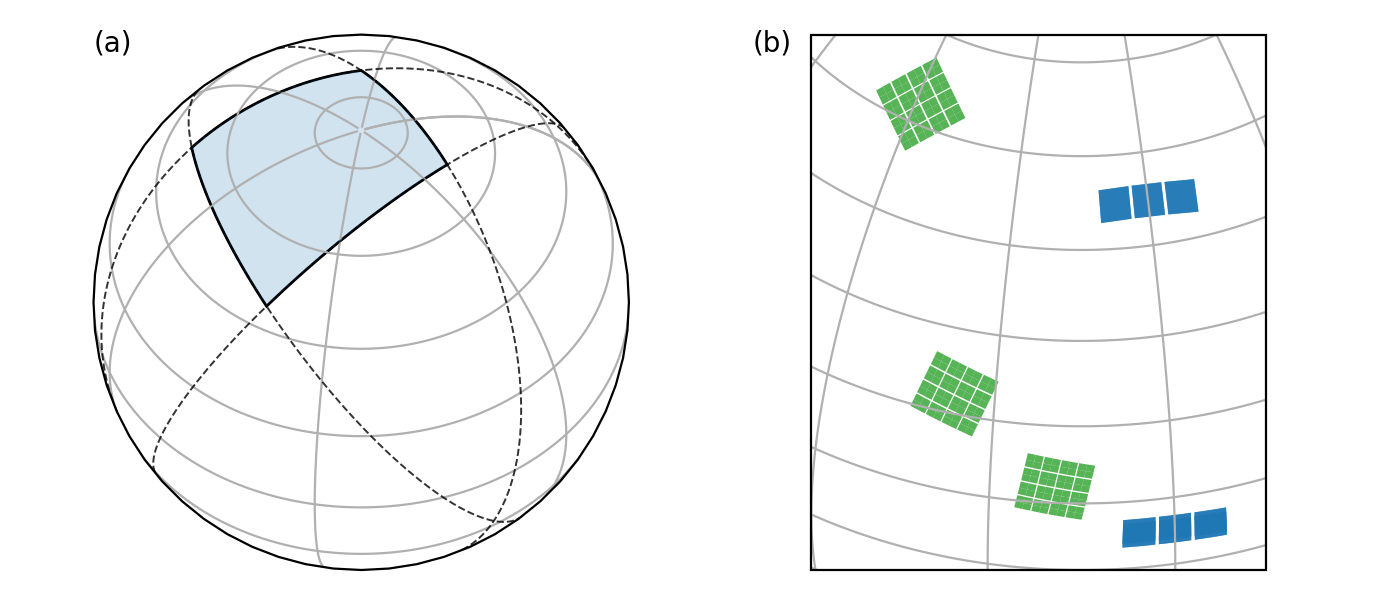}
\caption{Illustration of field-of-view (FoV) definitions and configurations. Panel a depicts the construction of a spherical polygon, where the FoV boundaries are defined by arcs of great circles on the celestial sphere. Panel b presents a zoomed-in example showing three actual ZTF FoVs composed of multiple individual spherical rectangles (green), along with two SPHEREx FoVs (blue) that shift toward lower latitudes as the survey proceeds. For scale, the image shown in Figure~\ref{fig:ztf_frame} corresponds to 1/64th of a single ZTF FoV.}
\label{fig:fovs}
\end{figure*}

The WISE/NEOWISE simulation focused on the initial ``Cryogenic'' phase of the WISE mission which began in January 2010, during which the spacecraft operated with cryogenic cooling across all four of its infrared bands (W1--W4). This phase represented the most observationally productive period of the mission, yielding over two million astrometric measurements of numbered asteroids submitted to the Minor Planet Center (MPC). During this period, WISE captured images of 1,491,686 unique patches of sky in all four bands simultaneously. Given the simultaneity of these exposures, \texttt{Kete} treats the four bands collectively for positional analysis, separating them only for flux estimation purposes. This simulation utilized a subset of the numbered asteroids for which diameters are either directly measured or reliably fitted; this constraint is necessary for the application of a thermophysical model. Here, we employed the Near-Earth Asteroid Thermal Model (NEATM; \citealt{Harris1998}) to predict the infrared brightness of each object. NEATM models asteroids as ideal blackbody, nonrotating spheres where the sunlit hemisphere is heated and re-radiates.
The resulting predictions and analysis are detailed in Section~\ref{flux}.

For these simulations, orbit propagation calculations were performed using an adaptive-precision algorithm. This method accepts a list of FoVs and a set of Cartesian states of asteroids or comets, and dynamically adjusts the propagation strategy to balance precision with performance. The algorithm begins by grouping the FoVs into batches based on time. Full $n$-body integrations are then performed between the median observation times of consecutive batches. By default, these batches of FoVs do not span more than 3 days of time, with $n$-body integration used to compute the state of the small bodies at the midpoint of the batch.

Within each batch, the motion of small bodies is then approximated using two-body dynamics, with the interval over which this approximation holds treated as a tunable parameter (which can be set to zero if strict $n$-body fidelity is required). This propagation step includes a correction for light delay due to the distance of the object from the observer. Following propagation, geometric visibility checks determine which objects fall within the instantaneous FoV of each image.

This hybrid approach of performing larger time step $n$-body integrations with short-term two-body approximations mitigates two major challenges: the computational cost of performing high-frequency $n$-body integrations for surveys with rapid cadences (e.g., WISE, which acquired multiple images per minute) and the accumulation of numerical error.

No matter the integration method used, every step of integration will introduce some unavoidable numerical noise (called Local Truncation Error). As more integration steps are performed, the Local Truncation Error will increasingly accumulate into a Global Truncation Error. To reduce the accumulation of numerical error it is important to keep the total number of integration steps to a minimum. The adaptive algorithm reduces the total number of integration steps by taking large integration steps between FoV batches, followed by a second short-term integration to the time of individual FoV observations. A demonstration of the benefit of this strategy is with the WISE dataset, where images were captured approximately every 12 seconds during operation. If $n$-body numerical integration were performed for each image, there would be approximately 200,000 integration steps per month, leading to rapid accumulation of error. By integrating in comparatively large one day steps to batch midpoints, and then integrating again within the batch for the time of each FoV, we reduce the rate of accumulation of error by a factor of 7,100 (the number of WISE images in a single day).

\section{Key Concepts}
\subsection{Geometry} \label{geometry}

\begin{figure*}[htb]
\centering
\includegraphics[width=0.9\textwidth]{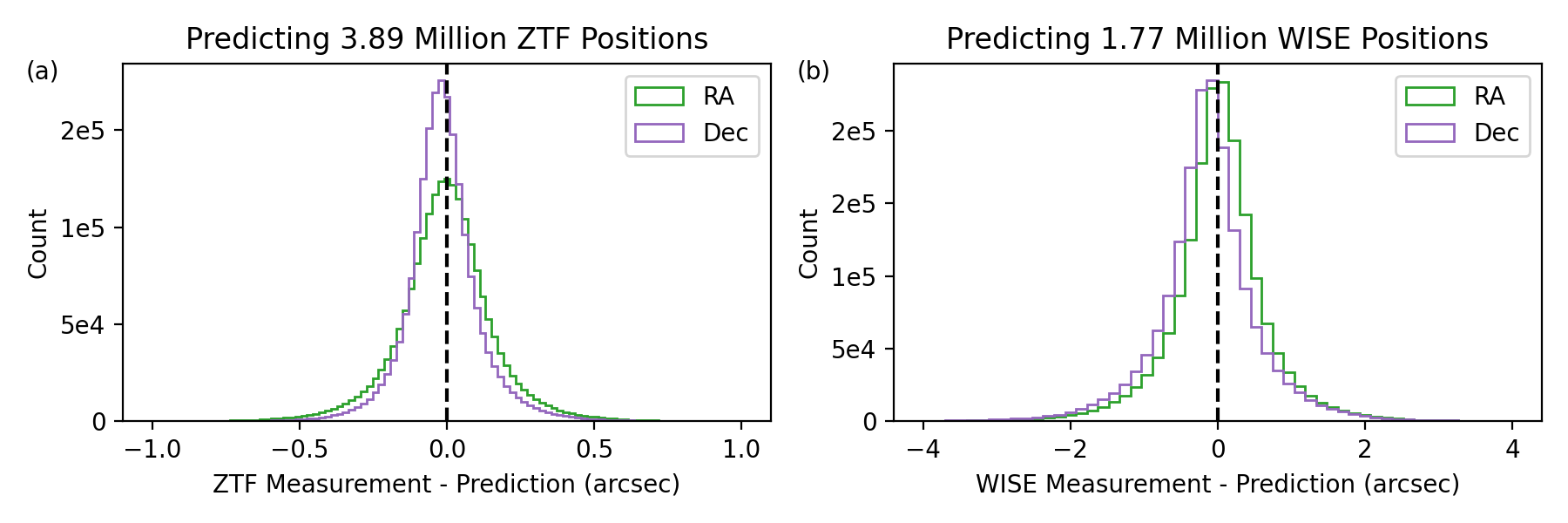}
\caption{Residuals between \texttt{Kete}'s predicted positions and astrometric positions reported to the MPC for all numbered asteroids detected during the simulation period. Panel a shows the residuals for ZTF; Panel b shows the residuals for WISE. Note that the WISE PSF Full Width Half Maximum (FWHM) is approximately 6\arcsec, making the residual errors larger than the ZTF residuals.}
\end{figure*}
\label{fig:residuals}

\begin{figure*}[htb]
\centering
\includegraphics[width=0.9\textwidth]{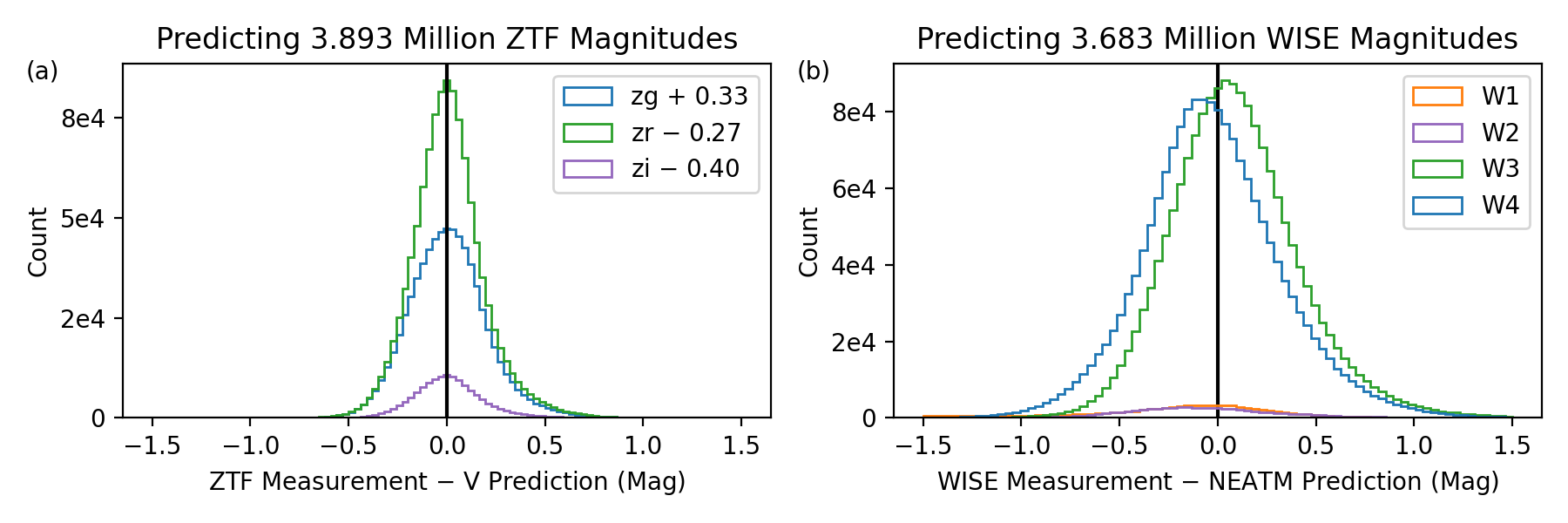}
\caption{Residuals between \texttt{Kete}'s predicted magnitudes and the magnitudes reported to the MPC for numbered asteroids captured by ZTF and WISE during the simulation periods. Panel a shows the ZTF residuals in $zg$, $zr$, and $zi$ bands, with an additional offset from predicted $V$-band magnitudes. Panel b shows the WISE equivalent, using the NEATM thermal model. Fewer counts are present in W1 and W2 as they were often below the detection threshold.}
\label{fig:flux_res}
\end{figure*}

To evaluate whether a Solar System object is observable from a given vantage point, we define FoVs as geometric patches on the celestial sphere. These FoVs can either be a simple cone where any object within a specific angular radius from the cone's center is considered visible, a spherical polygon (a closed figure on the surface of a sphere composed of great-circle arcs; see Fig.~\ref{fig:fovs}a), or a union of multiple spherical polygons (see Fig.~\ref{fig:fovs}b).

Spherical polygons are internally represented as collections of vectors which define the planes of the great circles. Each polygon thus corresponds to the intersection of multiple hemispherical regions (see Fig. \ref{fig:fovs}a). The normal vectors are oriented inward, such that the inclusion of an object can be efficiently verified via a series of dot products: if all dot products between the object’s position vector and the polygon’s plane normals are positive, the object resides within the FoV. This approach eliminates reliance on trigonometric functions and avoids singularities near the celestial poles, significantly improving computational efficiency. 

For instruments with multiple detectors, the overall FoV is defined as a collection of individual detector footprints. 
Fig.~\ref{fig:fovs}b demonstrates these multiple detector FoVs by showing actual footprints of images taken by SPHEREx and ZTF. Each FoV is associated with a specific observation time and the observer’s geocentric or heliocentric position. All observations are treated as instantaneous events (typically defined at the midpoint of the exposure duration); extended events can be approximated by constructing multiple individual observations (rightmost panel of Fig.~\ref{fig:ztf_wise_frame}).
This way of handling multiple detectors is also directly applicable to forthcoming missions, such as NEO Surveyor \citep{neos} and Roman \citep{Mosby2020}.

\subsection{Propagation} \label{prop}

\texttt{Kete} is designed for orbital propagation on the scale of decades to centuries. Accurate ephemerides for the majority of small bodies over these time scales are achieved by including the gravitational influence of the planets, the most massive main-belt asteroids, and corrections for general relativity due to the Sun and Jupiter. Additional perturbations, such as solar and Jovian oblateness, are also incorporated. The integration algorithm itself is implemented using a modified version of the \texttt{RADAU} integrator \citep{radau}, using some of the convergence criteria improvements inspired by the \texttt{IAS15} scheme \citep{ias15}. Further implementation details are provided in the Appendix \ref{app:radau}.

Calculating the positions of Solar System objects has a long history, and a number of software packages exist for this purpose. Although packages such as \texttt{Mercury} \citep{mercury} may be used to answer the same questions as \texttt{Kete}, their design is typically optimized for long-term (Myr--Gyr) dynamical studies of small-sized populations ($\sim$10$^5$ objects). \texttt{Kete} is optimized for high-throughput, short-to-medium-term analysis (Kyr) of much larger populations ($\sim$10$^9$ objects). These differences in goals lead to differences in design. For example, \texttt{Kete} leverages precomputed planetary ephemerides from \texttt{SPICE} kernel files (e.g., \texttt{DE440}; \citealt{de440}) to avoid redundant propagation of massive bodies during integration, resulting in significant performance gains. A natural downside of this approach is that the integration time span is restricted to the temporal range covered by the available kernel files.

\begin{figure*}[htb]
\centering
\includegraphics[width=0.8\textwidth]{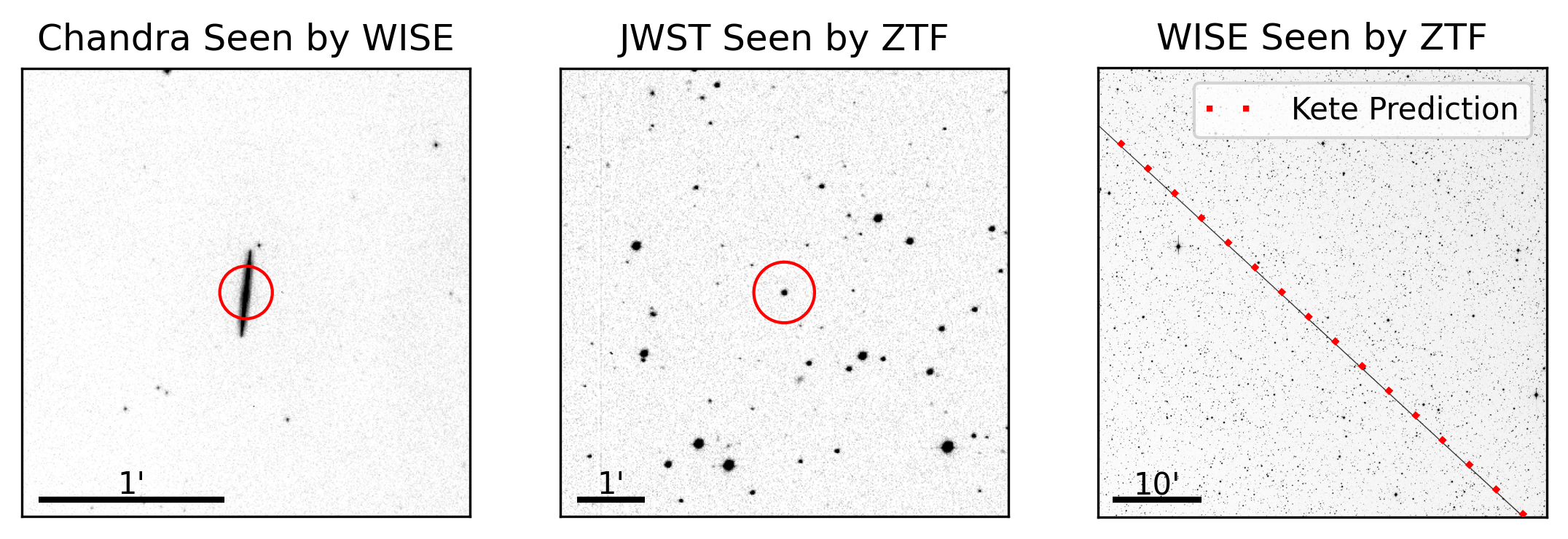}
\caption{Examples of mutual observability events predicted by \texttt{Kete}, in which one observatory views another. Such events occur between about 10 and 100 times per year for both WISE and ZTF. We were unable to find an instance of two observatories viewing each other.}
\label{fig:ztf_wise_frame}
\end{figure*}

The \texttt{SPICE} software has been commonly used for decades to store and read the position of Solar System objects from pre-computed files \citep{spice}. These pre-computed files are called \texttt{SPICE} kernel files, and are a compact binary representation of state vectors of Solar System objects. \texttt{Kete} employs a custom-built, natively multi-threaded \texttt{SPICE} kernel file reader to query the positions of massive bodies (planets and optionally other massive objects such as asteroids) during integration which does not rely on the existing \texttt{SPICE} code. This implementation enables fast numerical integrations as the planets themselves do not need to be propagated, so long as there is sufficient orbital information saved within the \texttt{SPICE} kernel files. \texttt{Kete} itself does not use or require the original \texttt{SPICE} software tools.

In order to validate the propagation accuracy of \texttt{Kete}, we compared its predicted ephemerides with actual observations submitted to the MPC by ZTF and WISE. \texttt{Kete} successfully recovered all 3.89 million ZTF and 1.77 million WISE MPC-submitted observations. Figure~\ref{fig:residuals} presents the RA and DEC residuals between \texttt{Kete}'s predictions and the corresponding MPC-reported positions, demonstrating that these residual errors match the known precision for ZTF and WISE MPC submissions \citep{nss-validation}. The slight offsets in the residuals appear to be due to very small time offsets on the scale of a few hundred milliseconds in the image metadata as queried from IRSA\footnote{IRSA (NASA/IPAC Infrared Science Archive) curates the science products of NASA's infrared and submillimeter missions (\url{https://irsa.ipac.caltech.edu/frontpage/}).}.

Beyond Solar System object prediction, \texttt{Kete} can also be used to identify artificial satellites. This can be seen in Figure~\ref{fig:ztf_wise_frame}, where the \texttt{SPICE} kernels were used to identify occasions when one observatory is observed by another. This functionality is not limited to observatories, as any object with an existing \texttt{SPICE} kernel may be used instead of $n$-body propagation to identify if the object is within an image.

\subsection{Flux Estimation}\label{flux}

\texttt{Kete} can predict apparent $V$-band magnitudes of asteroids using the H-G magnitude system \citep{Bowell1989}, consistent with the conventions adopted by the MPC and JPL Horizons. For cometary brightness estimates, \texttt{Kete} implements the magnitude models used by JPL Horizons\footnote{\url{https://ssd.jpl.nasa.gov/horizons/manual.html}} (see Appendix \ref{app:comet}). In the IR regime, it supports two thermophysical models of asteroids: the Near Earth Asteroid Thermal Model (NEATM; \citealt{neatm}) and the Fast Rotator Model (FRM; \citealt{frm}), enabling predictions of fluxes in bands such as those observed by WISE.

Figure~\ref{fig:flux_res} shows the residuals between the magnitudes predicted by \texttt{Kete} and those reported to the MPC, based on the simulations described in Section~\ref{prop}. Panel a shows residuals across the three ZTF bands, plotted relative to the predicted $V$-band magnitudes with intrinsic band offsets shown. Panel b displays analogous results for WISE, using NEATM-based flux predictions. The reduced number of W1 and W2 detections reflects the frequent non-detection of faint targets below the instrument’s sensitivity threshold.

Infrared flux predictions in \texttt{Kete} are computed using convex polygonal shapes, where any convex shape is supported (spherical geometry is the default). This faceted model was motivated by WISE requiring an additional temperature-dependent color correction to correctly estimate magnitudes (\citealt{wise}). In \texttt{Kete}, each facet emits blackbody radiation according to its temperature and orientation, allowing WISE-specific, temperature-dependent color corrections to be applied on a per-facet basis. The polygonal geometrical approach also governs the computation of reflected sunlight, where surfaces are modeled as ideal Lambertian scatterers with an additional phase correction based on the H-G system. 
When applied at a wavelength near 0.55~$\mu$m, the model yields results consistent with traditional $V$-band magnitude predictions. The contributions of both the blackbody emission and reflected light are then combined into the total apparent flux.

\section{Non-Survey Applications}

The following is a collection of applications of how \texttt{Kete} may be used. These applications demonstrate how the individual components of \texttt{Kete} may be composed together to answer questions. Additional examples are available in the online documentation\footnote{\url{https://dahlend.github.io/kete/}}.

\begin{figure*}[htb]
\centering
\includegraphics[width=0.8\textwidth]{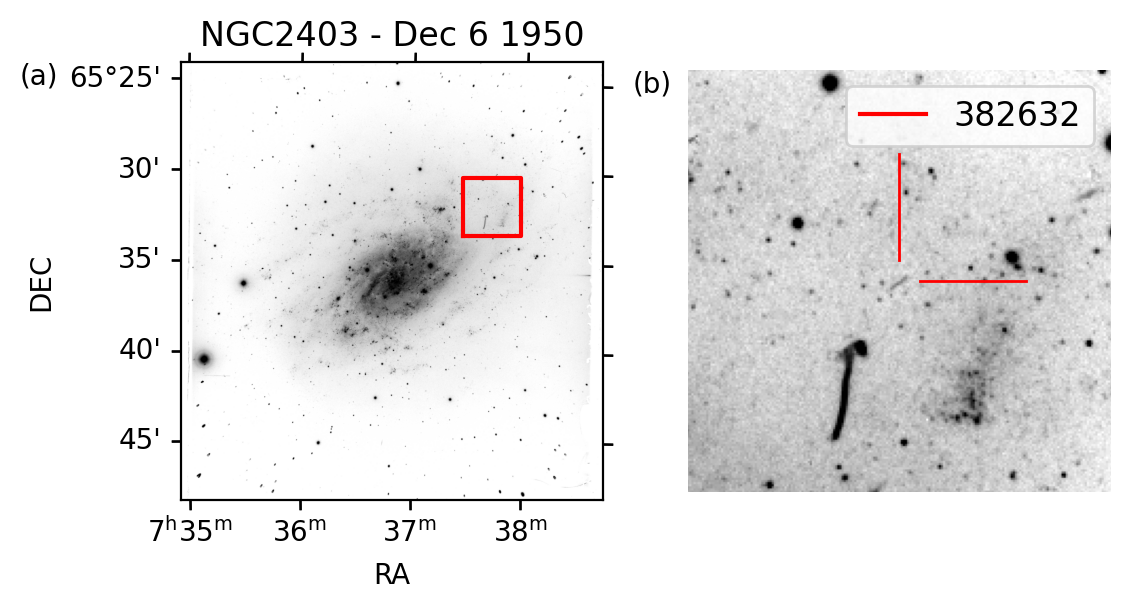}
\caption{Identification of an asteroid on a photographic glass plate taken at Palomar Mountain on 1950 December 6. Panel a shows the original target of the observation, NGC~2403, with a red box indicating the location of the asteroid. Panel b shows a zoomed-in view of the region, where the black arrow is the original hand-drawn marker for the asteroid, and red lines denote the position predicted by \texttt{Kete}. The asteroid (382632) is the streaked object.}
\label{fig:palomar}
\end{figure*}

\subsection{Pre-Discovery in Historic Archives}

In December of 1950, Rudolph Minkowski was observing variable stars from Palomar Mountain by taking 30-minute exposures onto photographic glass plates. Two of these plates, taken on December 6 and 8 and centered on NGC~2403, were later found to contain a faint asteroid streak. The streak was marked with a small hand-written arrow directly on the plates (see Fig.~\ref{fig:palomar}b), but no formal identification of the object was recorded at the time. Further details of these observations are documented in \cite{palomar1950}. The original glass plates were eventually archived at the Carnegie Science Observatories and later scanned for digital analysis.

We first conducted plate solving using the \texttt{Astometry.net} service\footnote{\url{https://astrometry.net/use.html}}. \texttt{Kete} was then employed to compute the position of the Hale telescope (observatory code 675) at the time of the observation and to define the corresponding FoV. We propagated the entire catalog of numbered asteroids backward to 1950, and their positions were evaluated against the FoV.


\begin{figure*}[htb]
\centering
\includegraphics[width=0.6\textwidth]{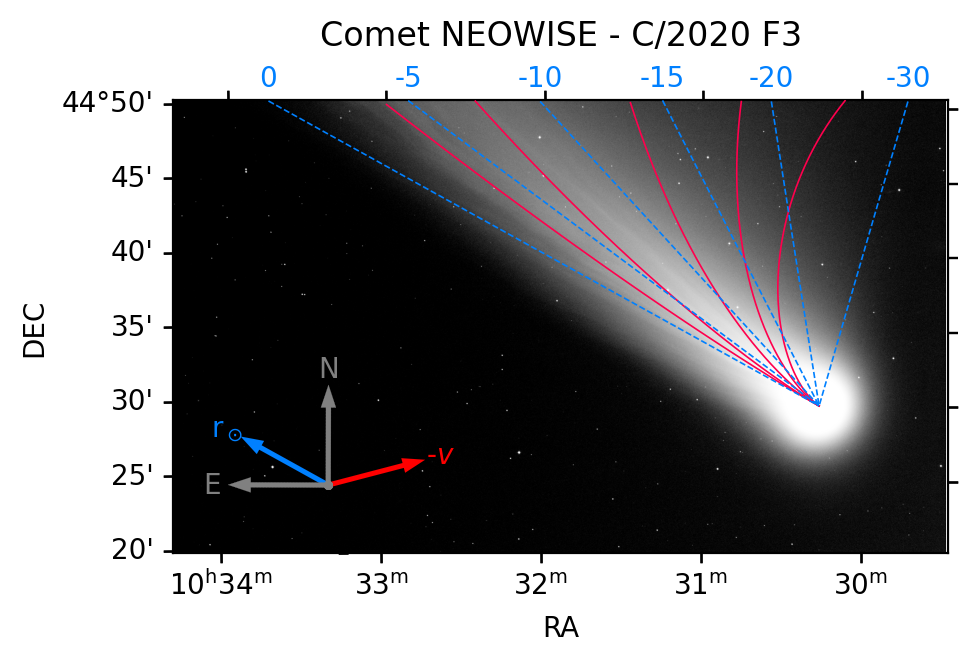}
\caption{Cropped image of Comet C/2020 F3 (NEOWISE) as seen from ZTF on 2020 July 23 at 03:51:15~UT. Overlaid lines illustrate an $n$-body implementation of the Finson-Probstein dust model \citep{Finson1968,Burns1979}. Blue dashed lines denote synchrones -- trajectories of dust particles released at specific times prior to observation, spanning a range of reduced solar radiation parameters ($\beta$). The labeled values correspond to the number of days. Red curves show syndynes -- paths of particles characterized by constant $\beta$ but released at varying times.}
\label{fig:com_neowise}
\end{figure*}

Among all known objects, only one object -- main-belt asteroid (382632) -- was found to lie within the plate’s FoV. The predicted position from \texttt{Kete} coincided precisely with the marked streak, confirming its identity. This asteroid was not officially discovered until 2002, also from Palomar Mountain Observatory, 52 years after the original observation.

\subsection{Dynamical Modeling of Cometary Dust}

\texttt{Kete} provides built-in support for simulating the dynamics of cometary dust, tracking the trajectories of particles ejected from the nucleus under the influence of solar gravity and non-gravitational forces. The model incorporates both Poynting-Robertson drag and solar radiation pressure acting on dust particles, allowing for simulations of dust particles in the tail of comets. 

Figure~\ref{fig:com_neowise} shows the application of this model to Comet C/2020 F3 (NEOWISE), as seen from ZTF $+$19 days from its perihelion passage. The overlaid trajectories represent an $n$-body implementation of the Finson-Probstein model \citep{Finson1968,Burns1979}. Dust in this computation was released with zero ejection velocity for simplicity, though \texttt{Kete} also supports user-defined ejection velocity and location of the source region in Cartesian coordinates. Synchrones (blue dashed lines) trace dust particles released at fixed epochs with a range of $\beta$ values, while syndynes (red solid curves) correspond to particles with fixed $\beta$ released across multiple times. Here, the reduced solar radiation parameter $\beta$ is defined as $\beta \propto (\rho_{\rm d} a)^{-1}$, where $\rho_{\rm d}$ is the dust density and $a$ is the particle radius. A detailed description of the force models is provided in Appendix \ref{app:nongrav_dust}.

\subsection{Evolution of Orbital Elements}

\begin{figure*}[htb]
\centering
\includegraphics[width=0.9\textwidth]{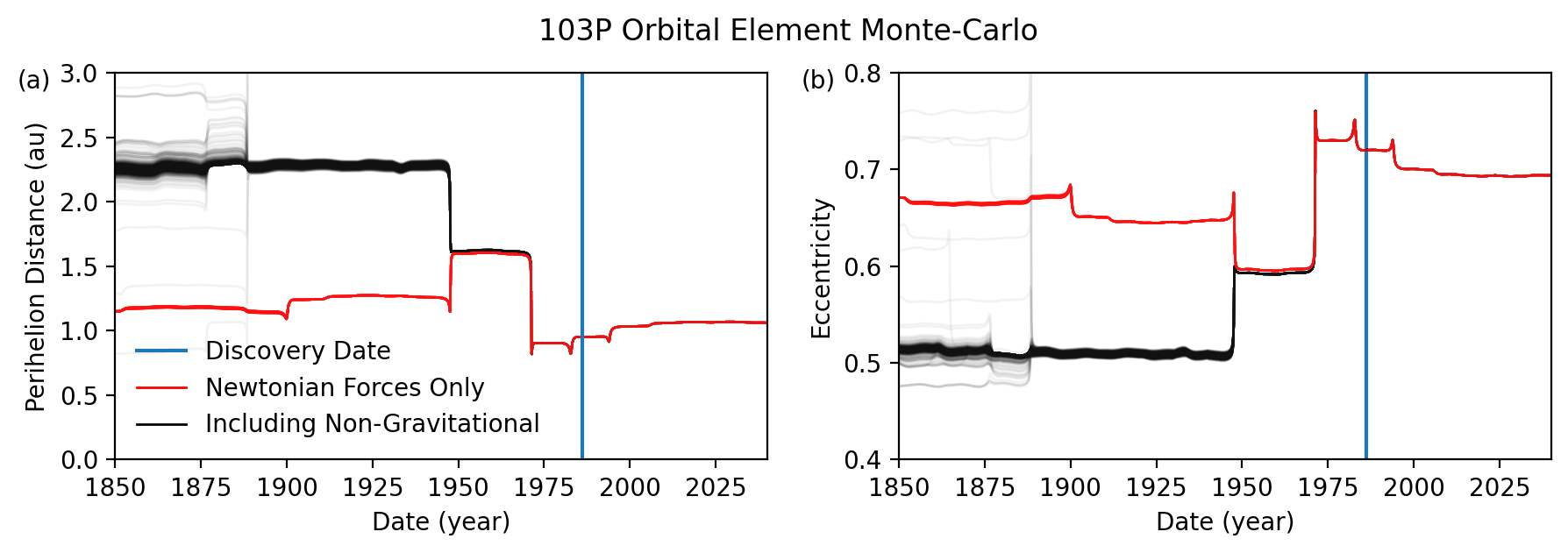}
\caption{
Monte Carlo simulation of 1,000 orbital realizations for Comet 103P/Hartley. Panels a and b present the temporal evolution of perihelion distance and eccentricity, respectively. Results are shown both with (black lines) and without (red lines) the inclusion of non-gravitational forces. The red-line solution, excluding non-gravitational effects, is consistent with the orbital history derived by \citet{Ip2016} using \texttt{Mercury}. The blue vertical line marks the comet’s discovery date (1986 March 15, by Hartley).
}
\label{figure:103P}
\end{figure*}

Comet 103P/Hartley, discovered on 1986 March 15 by Hartley\footnote{\url{https://ssd.jpl.nasa.gov/tools/sbdb_lookup.html#/?sstr=103P}}, is a hyperactive Jupiter-family comet and the smallest comet nucleus visited by a spacecraft to date \citep{Harker2018}. To investigate its orbital evolution, we used \texttt{Kete} to draw 1,000 correlated samples from the covariance matrix of the orbit fit provided by JPL Horizons. Each sample was constructed into a Cartesian state vector and propagated backward to the year 1850, including the effects of non-gravitational forces. Orbital elements were computed at 30-day intervals for each trajectory. 

Figure~\ref{figure:103P} presents the resulting evolution of perihelion distance and eccentricity. Results both with and without incorporating non-gravitational forces were provided for comparison. The date of discovery was shortly after a noticeable decrease in perihelion distance to below 1~au. This transition is associated with a close encounter with Jupiter (centered on 1971 April 29th at a distance of 0.0856~au), which significantly modified the orbit and improved the comet’s observational accessibility from Earth.

\begin{figure*}[htb]
\centering
\includegraphics[width=0.83\textwidth]{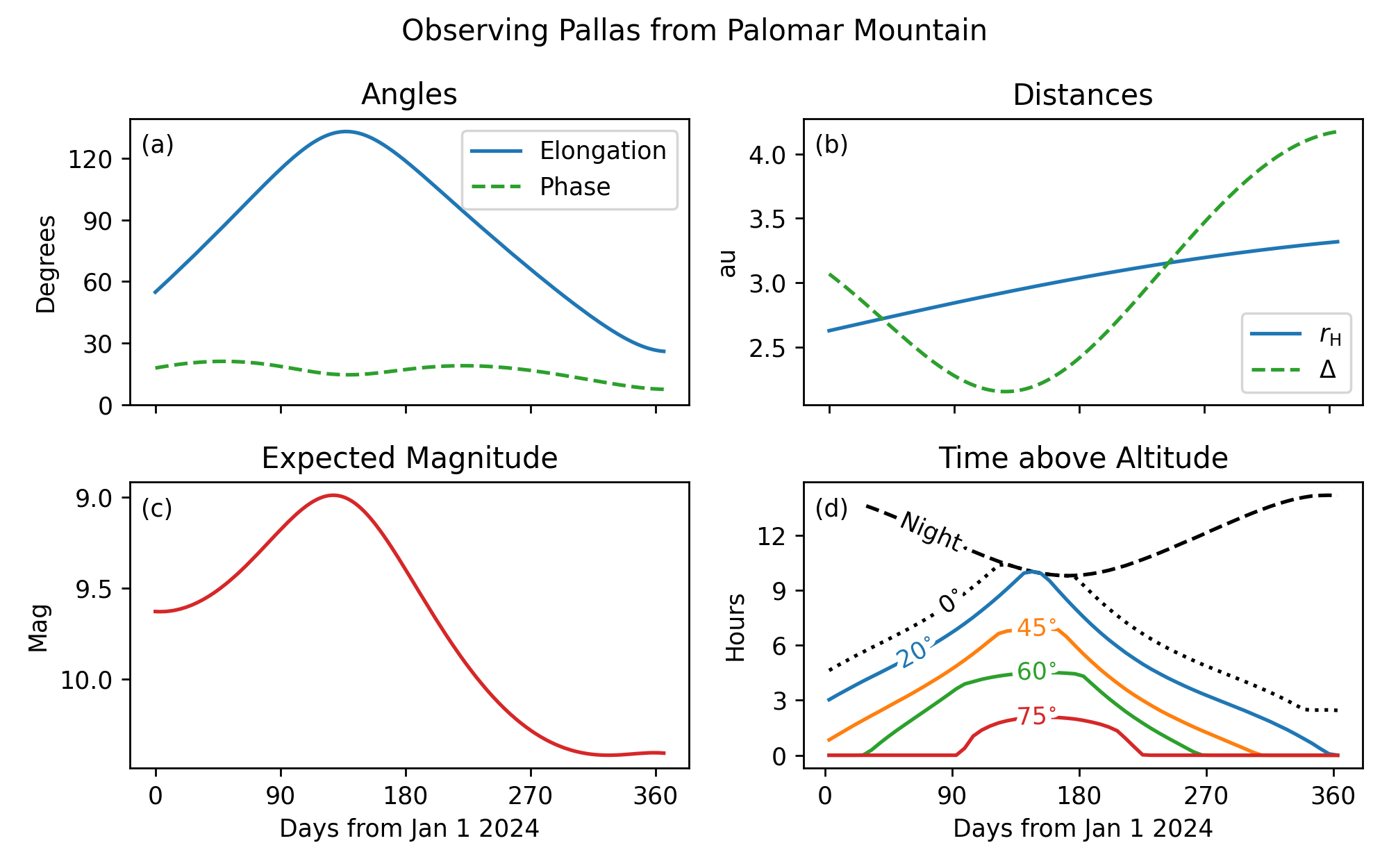}
\caption{Example observation planning output from \texttt{Kete}, with all orbit and magnitude calculations performed locally. Panels a--c displays the changes in geometry and magnitude of (2)~Pallas over the course of the year, where $r_{\text{H}}$ and $\Delta$ are the heliocentric and geometric distance in au, respectively. Panel d shows the number of hours per night during which (2)~Pallas is visible above a specified altitude from the Palomar Mountain Observatory, showing that it remains near zenith for several hours during certain months.}
\label{fig:obs_plan}
\end{figure*}

\subsection{Observation Planning}
\texttt{Kete} also facilitates observation planning for Solar System bodies by computing sky positions and observability windows from any specified location. The functionality supports both MPC observatory codes and user-defined coordinates using the \texttt{WGS84} geodetic reference coordinate system. Initial orbital elements used for propagation were retrieved from either JPL Horizons or the MPC. 
By accounting for site-specific observing geometry, including altitude, airmass, and duration of visibility in hours, \texttt{Kete} enables users to identify the optimal observation windows for individual objects. Figure~\ref{fig:obs_plan} illustrates a use case for (2)~Pallas, showing its changing geometry and visibility from Palomar Mountain over the course of 2024.

\section{Summary} \label{sec:summary}

We present \texttt{Kete}, a high-performance, new open-source software framework for identifying and modeling the visibility of known asteroids and comets in both archival and real-time astronomical survey data. Designed to address big data from modern all-sky surveys, \texttt{Kete} combines fast geometric field-of-view filtering, accurate orbital propagation over kilo-year timescales, and photometric prediction in both optical and infrared regimes.

The framework is implemented in Python with a compiled Rust backend for computational efficiency, enabling large-scale simulations of billions of object-image intersections on commodity hardware. Core functionalities include support for arbitrary detector layouts, multi-core support for \texttt{SPICE} kernels, magnitude predictions via the H-G (optical) and NEATM/FRM thermophysical models (IR), and dynamical modeling of cometary dust using an $n$-body Finson–Probstein formalism with non-gravitational forces. These capabilities are organized into modular components that can be flexibly combined for applications ranging from observation planning and survey simulation to historic pre-discovery and orbital evolution studies.

We demonstrated \texttt{Kete}'s utility through two full-year simulations: one for ZTF (ground-based, optical) and another for the cryogenic phase of WISE (space-based, IR). These simulations predicted the positions and magnitudes of hundreds of thousands of numbered asteroids across more than 11 million images, achieving full recovery of all known numbered asteroid detections submitted to the MPC during the corresponding periods. Residual astrometric and photometric errors were consistent with the known precision of each survey.

Beyond survey applications, we applied \texttt{Kete} to a range of representative use cases: recovering pre-discovery detections on historic photographic plates taken in 1950, modeling cometary dust tail morphology, characterizing the orbital evolution of comet 103P/Hartley, and planning observations of asteroid (2)~Pallas. These examples highlight \texttt{Kete}'s versatility and readiness for integration into next-generation survey analysis and science workflows. 

Future plans include extending gravitational models to support more advanced dust models, along with more complex optical and thermal models. 

\begin{acknowledgements}
We would like to acknowledge the entire Near Earth Object Surveyor team (past and present), many of whom indirectly contributed to this work by using \texttt{Kete} and its earlier versions for their own work, providing valuable feedback. We also acknowledge the public archives which this library uses frequently, IPAC IRSA, JPL Horizons \& NAIF, and the Minor Planet Center. We would like to thank Carnegie Science Observatories and specifically Archivist Kit Whitten for providing the Palomar historic plate for analysis.
\\
This publication makes use of data products from the Wide-Field Infrared Survey Explorer, which is a joint project of the University of California, Los Angeles, and the Jet Propulsion Laboratory/California Institute of Technology, funded by the National Aeronautics and Space Administration. This publication also makes use of data products from NEOWISE, which is a project of the Jet Propulsion Laboratory/California Institute of Technology, funded by the Planetary Science Division of the National Aeronautics and Space Administration. This research has made use of the NASA/IPAC Infrared Science Archive, which is funded by the National Aeronautics and Space Administration and operated by the California Institute of Technology. This publication also makes use of software and data products from the NEO Surveyor, which is a joint project of the University of California, Los Angeles and the Jet Propulsion Laboratory/California Institute of Technology, funded by the National Aeronautics and Space Administration.
\end{acknowledgements}

\bibliographystyle{aa}
\bibliography{refs}

\appendix

\section{Integrator Details}

\label{app:a}
The primary $n$-body integration method used in \texttt{Kete} relies on pre-computed \texttt{SPICE} kernel files for all massive objects. Using this method, integration can only be performed on massless test particles. By default \texttt{Kete} automatically downloads and caches the \texttt{DE440} \texttt{SPICE} kernel \citep{de440} which contains the barycenter of the planets, along with the Earth and Moon positions.  New massive objects can be registered with the numerical integrator, however \texttt{SPICE} kernels for these objects must be loaded. The default list of registered massive objects are the planets, the Sun, the Moon, and optionally the largest main belt asteroids. The masses of the planets and the Sun used during integration were taken from \texttt{DE440}. The total acceleration each non-massive object experiences is a sum of the respective correction terms along with Newtonian gravitational acceleration from all registered masses:
\begin{align}
\vec{a}_\text{total} = \vec{a}_{\text{rp}} +  \vec{a}_{\text{com}} + \sum_{\text{m}} \bigg(\frac{\mu_\text{m}\,\vec{r}}{|\vec{r}|^3} + \vec{a}_{\text{rel},\text{m}} + \vec{a}_{\text{J2,m}} \bigg)
\end{align}
Where m indicates the central mass that causes the force, $\vec{r}$ is the relative position between the two masses, $\mu_\text{m}$ is the standard gravitational parameter of the central body m in AU$^3$ / day$^2$. The acceleration terms $a_{\text{rel,m}}$ and $a_{J2\text{,m}}$ are a general relativistic correction and non-spherical correction terms which are applied to the Sun, Jupiter, and Earth. Solar radiation pressure ($\vec{a}_\text{rp}$) and the Cometary non-gravitational ($\vec{a}_\text{com}$) terms are optional and not included by default.

\subsection{Units \& Frames}

Throughout \texttt{Kete}, the default units of distance are AU and units of time are Barycentric Dynamical Time (TDB) scaled days. \texttt{Kete} provides conversion tools between different time scales and formats. The default coordinate frames are defined to match the \texttt{SPICE} implementation of the Equatorial J2000, Ecliptic J2000, and Galactic coordinate frames, where \texttt{SPICE} considers the equatorial implementation equivalent to the modern ICRF reference frame \citep{ICRF}. In addition to this, \texttt{Kete} supports non-inertial frames, which enables accurate conversion of state vectors from Earth's surface to state vectors in the Solar System. This includes support for the high-precision Earth frame orientation files available through the JPL NAIF website\footnote{Navigation and Ancillary Information Facility (NAIF): \url{https://naif.jpl.nasa.gov/pub/naif/generic_kernels/pck/}}, as well as an implementation of a precession and nutation model of Earth's orientation \citep{earthprecession}.

\subsection{Relativistic Corrections}\label{app:gr}

In addition to Newtonian gravity, \texttt{Kete} adds a correction term for relativistic effects from Equation 2.5 in \cite{damour:1985}, where the non-Newtonian first-order correction is reproduced here:
\begin{align}
\begin{split}
\vec{a}_\text{rel,m} = \frac{\mu_\text{m}}{c^2|\vec{r}|^3}\bigg[&\bigg(\frac{\mu_\text{m}}{|\vec{r}|}(4 + 2\epsilon) - |\vec{v}|^2(1 + 3\epsilon) + \frac{3\epsilon}{2}\bigg(\frac{\vec{r}\cdot\vec{v}}{|\vec{r}|}\bigg)^2\bigg) \,\vec{r}\\
&+ (4 - 2\epsilon)\big(\vec{r}\cdot\vec{v}\big)\,\vec{v}
 \bigg]
\end{split}
\end{align}

\noindent where $\vec{v}$ is the relative velocity, $c$ is the speed of light, and the reduced mass $\epsilon = mm' / (m + m')^2$. In \texttt{Kete}, the mass of individual particles is $m'=0$, making the reduced mass $\epsilon=0$. The equation above then simplifies to the form seen on page 820 in Asteroids IV \citep{asteroidsIV}:
\begin{align}
\vec{a}_\text{rel,m} = \frac{\mu_\text{m}}{c^2|\vec{r}|^3}\bigg[\bigg(\frac{4\mu_\text{m}}{|\vec{r}|} - |\vec{v}|^2 \bigg)\,\vec{r} + 4(\vec{r}\cdot\vec{v})\,\vec{v}
 \bigg]
\end{align}

\noindent This correction term is added for accelerations due to the Sun, Jupiter, and Earth. This is an approximation, and a more precise general form is discussed in \cite{moyer:2003}, specifically see their Section 4.4.1.

\subsection{Non-Spherical J$_2$ Corrections}\label{app:j2}
Bodies in the Solar System may be approximated as oblate spheres. A spherical harmonic expansion can be used to approximate the gravitational forces of these oblate spheres. The first correction for this non-sphericity is denoted J$_2$, which defines how much the sphere ``bulges" at its equator. Below are the gravitational acceleration terms \texttt{Kete} uses to account for this correction.
\begin{align}
\begin{split}
  \vec{a}_{\text{J2,m}} &= \psi_\text{m}\,\bigg[\bigg(1-5\frac{r_z^2}{|\vec{r}|^2}\bigg)\,r_x,\,\,
   \bigg(1-5\frac{r_z^2}{|\vec{r}|^2}\bigg)\,r_y,\,\,
   \bigg(3-5\frac{r_z^2}{|\vec{r}|^2}\bigg)\,r_z\bigg]
    \\
  &\text{where} \qquad \psi_\text{m} \equiv \frac{3}{2}\frac{\text{J}_{2,{\rm m}} \,\mu_\text{m}\, \rho_\text{m}^2 }{|\vec{r}|^5}
\end{split}
\end{align}

\noindent where $\rho_\text{m}$ is the radius of the central body in AU, $\mu_\text{m}$ is the standard gravitational parameter of the central body m in AU$^3 / \text{day}^2$, and the relative position vector $\vec{r} \equiv [r_x, r_y, r_z]$ in units of AU. The position vector is defined in respect to the equatorial frame for the planet, where the $z$ axis is perpendicular to the equatorial plane of the planet. Note that in the acceleration equation, the $z$-component is different from the $x$ and $y$. \texttt{Kete} applies this correction term for Jupiter and the Sun, and this term is 0 for all other planets by default.

\subsection{Non-gravitational Forces - Cometary Dust} \label{app:nongrav_dust}
\texttt{Kete} supports two non-gravitational force models, the first is solar radiation pressure, including Poynting-Robertson drag from \cite{radiationpressure}:
\begin{align}
\vec{a}_\text{rp} = \frac{L_0\, A\, Q_{pr}}{|\vec{r}|^2 c \,\text{m}_\text{dust}} \bigg[\bigg(1 - \frac{\vec{v}}{c}\bigg)\, \vec{r} - \vec{v} / c \bigg]
\end{align}

\noindent where $L_0$ is the luminosity of the Sun in $\text{W}/\text{m}^2$ at 1 AU, $A$ is the effective cross sectional area of the dust in m$^2$, $Q_{pr}$ is a unit-less scattering coefficient, and $\text{m}_\text{dust}$ is the dust mass in kg. Defining the term $\beta$, we arrive at the more commonly seen form:
\begin{align}
\begin{split}
\vec{a}_\text{rp} &= \frac{\beta \,\mu_\text{sun}}{|\vec{r}|^2} \bigg[\bigg(1 - \frac{\vec{v}}{c}\bigg)\, \vec{r} - \vec{v} / c \bigg] \\
&\text{where}\qquad \beta \equiv \frac{L_0 \,A \,Q_{pr}}{c \,m_{\text{dust}} \,\mu_\text{sun}}
\end{split}
\end{align}

\noindent  \texttt{Kete} only accepts $\beta$ as an input for non-gravitational dust modeling, leaving the computation of $\beta$ from mass, density, and effective cross-sectional area up to the user.

\subsection{Non-gravitational Forces - JPL Comet Model} \label{app:nongrav}
The second non-gravitational force model supported is the model from \cite{cometsII}, specifically pages 138--139:
\begin{align}
\begin{split}
\vec{a_\text{com}} &= A_1 g(r) \vec{r} + A_2 g(r) \vec{t} + A_3 g(r) \vec{n}\\
&\text{where}\qquad g(r) \equiv \alpha\, \bigg(\frac{r}{r_0}\bigg) ^ {-m} \bigg[1 + \bigg(\frac{r}{r_0}\bigg) ^ n\bigg] ^ {-k}
\end{split}
\end{align}
\noindent where the vectors $\vec{r}$, $\vec{t}$, and $\vec{n}$ represent the radial, transverse, and normal unit vectors of the motion of the comet. $A_i$ ($i$ = 1, 2, 3), $\alpha$, $r_0$, $n$, $m$, and $k$ terms are all free parameters of the model, which are all unitless except for $r_0$ which is a distance in AU. This is the model which JPL Horizons uses, with defaults for comets defined as $\alpha=0.1112620426$, $r_0=2.808$ AU, $m=2.15$, $n=5.093$, and $k=4.6142$ unless otherwise specified. Non-gravitational forces for asteroids use the same model, but different default parameters, $\alpha=1.0$, $r_0=1.0$ AU, $m= 2.0$, $n=1.0$, $k=0.0$, which causes the equation to be simplified to $g(r)=1/r^2$.

This model has the added complication that forces on the comet might be evaluated due to its position at a time in the past. This is due to the current activity being influenced by thermal lag. In order to account for this, this model accepts an additional $dt$ term, which uses a two-body approximation to estimate the comet position $dt$ days in the past, then evaluates the current acceleration from that previous position. This matches the implementation as defined in the JPL Horizons online documentation\footnote{\#9. Visual magnitude \& surface brightness at https://ssd.jpl.nasa.gov/horizons/manual.html}.

\subsection{RADAU Modifications} \label{app:radau}
With the improved convergence testing introduced by \texttt{IAS15} \citep{ias15}, we found that we could entirely remove the predictor step of the \texttt{RADAU} integrator \citep{radau}, and use the convergence testing to iterate more during the corrector step if required. This change increased performance significantly (approximately double), and we found that except during close encounters, does not introduce accuracy issues. We demonstrate in the main text that this integrator works well for several decades, however we believe this integrator should not be used for any long-term (Gyr) simulations.

The convergence test parameters that are defined in \texttt{IAS15} \citep{ias15} are still used, but have been tuned in our implementation to achieve computational performance at the cost of precision. The precision goals during this tuning were to achieve less than approximately 70~km difference over a 10-year orbit in computed positions against JPL Horizons computed values. This value is approximately equivalent to a maximum residual error 100 mas when an object is viewed at 1 au distance.

\section{Comet Magnitudes}
\label{app:comet}

Comet magnitudes in \texttt{Kete} use the default model as defined by the JPL Horizons service:
\begin{align}
T_{\rm mag} &= M_1 + 5\log(\Delta) + K_1\log(r_{\rm H})\\
N_{\rm mag} &= M_2 + 5\log(\Delta) + K_2\log(r_{\rm H}) + \frac{0.035}{1^\circ} \Psi
\end{align}

\noindent where $T_{\rm mag}$ is the total magnitude of both the nucleus and extended signal. $N_{mag}$ is the estimated magnitude of only the nucleus. $M_{1/2}$ and $K_{1/2}$ are model parameters that \texttt{Kete} can query from JPL Horizons. $\Delta$ and $r_{\rm H}$ are the geocentric and heliocentric distances of the comet in au, respectively. $\Psi$ is the phase angle (angle between the Sun--Comet--Observer) in degrees, and constant 0.035 is the JPL horizon's default phase correction coefficient.

%
%

\end{document}